\documentclass[11pt,twoside]{article}

\usepackage{asp2006_hotcool}
\usepackage{epsf}
\usepackage{lscape}
\usepackage{graphicx}

\begin{document}
\title{The Physical Properties of Red Supergiants} 
\author{Emily M. Levesque$^1$}
\affil{Institute for Astronomy, University of Hawaii, 2680 Woodlawn Dr., Honolulu, HI 96822}
\footnotetext[1]{Predoctoral Fellow, Smithsonian Astrophysical Observatory}

\begin{abstract}
Red supergiants (RSGs) are a He-burning phase in the evolution of moderately massive stars (10-25M$_{\odot}$). For many years, the assumed physical properties of these stars placed them at odds with the predictions of evolutionary theory. We have recently determined new effective temperatures and luminosities for the RSG populations of galaxies with a factor of $\sim$8 range in metallicity, including the Milky Way, the Magellanic Clouds, and M31. We find that these new physical properties greatly improve the agreement between the RSGs and the evolutionary tracks, although there are still notable difficulties with modeling the physical properties of RSGs at low metallicity. We have also examined several unusual RSGs, including VY CMa in the Milky Way, WOH G64 in the LMC, and a sample of four RSGs in the Magellanic Clouds that show considerable variations in their physical parameters, most notably their effective temperatures. For all of these stars we reexamine their placement on the H-R diagram, where they have appeared to occupy the ``forbidden" region to the right of the Hayashi track. We have updated current understanding of the physical properties of VY CMa and WOH G64; in the case of the unusual Magellanic Cloud variables, we conclude that these stars are undergoing an unstable evolutionary phase not previously associated with RSGs. 
\end{abstract}

\section{Introduction}
Red supergiants (RSGs) are an important, He-burning phase in the evolution of moderate massive (10-25M$_{\odot}$) stars. Until recently, the location of these stars on the H-R diagram did not agree with the predictions of stellar evolutionary theory, appearing to be too cool and too luminous to coincide with the positions of the evolutionary tracks. This placed them at odds with the restrictions imposed by the Hayashi limit, which denotes the largest radius a star of a given mass can have while still remaining in hydrostatic equilibrium (Hayashi \& Hoshi 1961). Stars with effective temperatures that are cooler than the evolutionary tracks allow occupy a ``forbidden" region to the right of this limit on the H-R diagram.

There were several possible explanations for why RSGs showed this systematic disagreement with theoretical predictions. For one, these stars present significant challenges to evolutionary theory. The velocities of the convective layers are nearly sonic and even super-sonic in the atmospheric layers, giving rise to shocks \cite{F2002} and invalidating mixing-length assumptions. There are also uncertainties in our knowledge of RSG molecular opacities. Finally, the highly-extended atmospheres of these stars differ from the plane-parallel geometry assumption of the evolutionary models; spherical models would be more appropriate. As a result of these difficulties, the fault for this disagreement between observed and theoretical properties of RSGs could well have been with the the placement of the evolutionary tracks on the H-R diagram.

However, another possibility was that the ``observed" location of RSGs in the H-R diagram was incorrect, a result of inaccurate determinations of their effective temperatures. Past attempts at determining effective temperature scales for RSGs have been forced to make several assumptions that limit their accuracy. For example, Humphreys \& McElroy (1984) determine effective temperature (T$_{\rm eff}$) by assuming a blackbody continuum and using broadband colors to assign T$_{\rm eff}$ based on the few nearly RSGs with measured diameters (Lee 1970, Johnson 1964, 1966). However, in addition to problems with the blackbody assumption, increased line blanketing at the low surface gravities of RSGs severely affects the ($B-V$) colors (for more discussion of this effect see Massey 1998). A more recent T$_{\rm eff}$ scale from Massey \& Olsen (2003) adopts the Dyck et al.\ (1996) scale for red $giants$, determined through interferometric data for these stars, and shifts the temperatures down by 400 K; again, such an approach is merely a rough estimate of the T$_{\rm eff}$ scale expected for red supergiants.

A better approach for determining T$_{\rm eff}$ would be to make use of the TiO bands that dominate the optical spectra of M-type RSGs; these lines are used to determine spectral subtype and are quite temperature-sensitive. However, until recently, atmospheric models for RSGs have not included accurate treatments of molecular opacities. Fortunately, the new generation of MARCS models (Gustaffson et al.\ 1975, 2003, 2008; Plez et al.\ 1992, Plez 2003) now includes a much improved treatment of molecular opacity. With these models, it is now possible to construct a much more accurate T$_{\rm eff}$ scale for RSGs.

\section{Red Supergiants in the Milky Way}
In Levesque et al.\ (2005) we present moderate-resolution ($\sim$ 5\AA) optical spectrophotometry of 74 RSGs in the Milky Way. We redetermined the spectral types of these stars, and fit our spectra with the MARCS stellar atmosphere models. The goodness of our fits was primarily based on the strengths of the TiO bands used for spectral type classification - for an example of our fitting see Figure 1. From these fits we were able to determine T$_{\rm eff}$, for all our stars and construct a new T$_{\rm eff}$ scale for Galactic RSGs.  In addition, we used the MARCS models to determine a relation between T$_{\rm eff}$ and RSG ($V-K$)$_0$ colors, obtaining comparable parameters from both this method and our spectral fitting.

While fitting our RSG spectra, we noticed two things: first, that the extinctions we had to adopt to obtain satisfactory fits were usually in excess of what was reported for OB stars in the same associations, and second, that there was considerably more stellar NUV flux in these cases (see example in Figure 1). We were later able to confirm that this excess flux in the NUV was well-correlated with the amount of excess extinction, dust production rate, and the star's bolometric luminosity (Massey et al.\ 2005). Based on these findings, we concluded that the excess NUV flux was indicative of the presence of circumstellar dust, most likely the result of preferential scattering of blue light by off-axis parts of the dust shell. Finally, Massey et al.\ (2005) states that this circumstellar material may not follow the standard $R_V = 3.1$ reddening of Cardelli et al.\ (1989); determination of the grain scattering properties of this circumstellar material using Spitzer data  is underway.

We were also able to use data from the MARCS models to calculate the T$_{\rm eff}$-sensitive bolometric corrections for these stars at both $V$ and $K$, allowing us to determine the bolometric luminosities (M$_{\rm bol}$ for these stars and place them on the H-R diagram. These newly derived physical properties resulted in an important leftward and downward shift of RSGs on the H-R diagram - there is now excellent agreement between Galactic RSGs and the predictions of the Geneva evolutionary tracks (Meynet \& Maeder 2003). This suggests that, in general, stellar evolutionary theory does a good job of accurately reproducing this stage of massive star evolution in the Milky Way. However, occasionally there are particularly unusual RSGs (in the Milky Way and, as we later found, in the Magellanic Clouds and M31 as well) that require a closer look.

\begin{figure}
\begin{center}
\includegraphics[scale=0.6]{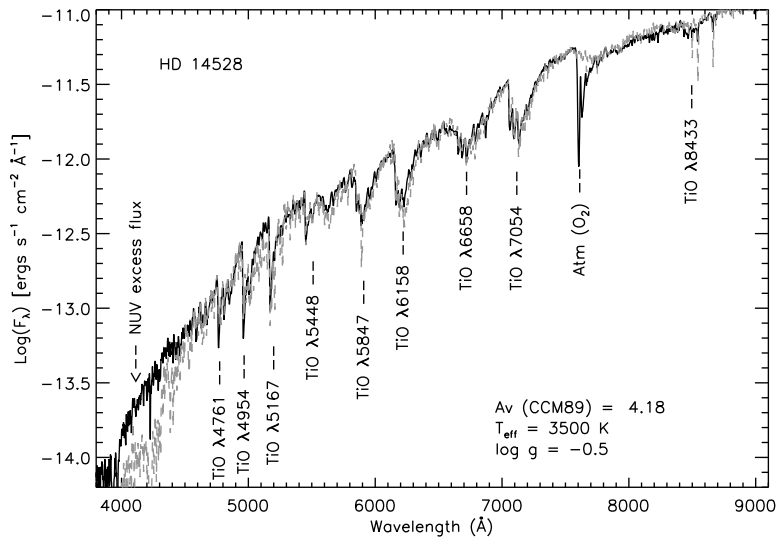}
\end{center}
\caption{Example of our spectra fitting; a spectrum of Galactic RSG HD 14528 (black solid line) overplotted with the best-fit MARCS stellar atmosphere model (gray dotted line); there is good agreement with the depths of all the TiO bands. The strong feature at 7600\AA\ is the telluric A band. The effective temperature, log g, and $A_V$ are given for the model fit. The excess NUV flux in the observed SED as compared to the model SED is noted.}
\end{figure}

\subsection{VY Canis Majoris}
VY Canis Majoris is a peculiar M supergiant in the Milky Way, with several physical characteristics that set it apart as extreme as compared to RSGs. It has a large IR excess, indicative of a circumstellar dust shell or disk being heated by the star (Herbig 1970). Such a dust shell, an asymmetric dust reflection nebula extending 8"-10" from the star, has since been observed and studied in great detail (Monnier et al.\ 1999; Smith et al.\ 2001; Smith 2004; Humphreys et al.\ 2005). VY CMa also has a particularly high mass-loss rate for a RSG; about 2 $\times 10^{-4} M_{\odot}$ yr$^{-1}$ (Danchi et al.\ 1994). Until recently, this star was assumed to have extreme physical properties as compared to other Galactic RSGs (Le Sidaner \& Le Bertre 1996; Smith et al.\ 2001; Monnier et al.\ 2004; Humphreys et al.\ 2005); its assumed T$_{\rm eff}$ for many years was extraordinarily cool (e.g., 2800 K in Le Sidaner \& Le Bertre 1996, 3000 K in Smith et al.\ 2001), based on the star's spectral type and prior T$_{\rm eff}$ scales such as that of Dyck et al. (1974). This placed VY CMa in the forbidden region to the right of the Hayashi track on the H-R diagram and resulting in calculation of a very large radius for this star.

Massey et al.\ (2006a) used the MARCS models to redetermine VY CMa's temperature spectroscopically, finding a satisfactory fit of T$_{\rm eff} = 3650$ K based on the redder TiO bands (6158\AA, 6658\AA, 7054\AA); focusing on the bluer bands (4761\AA, 4954\AA, 5167\AA) gave a cooler (and slightly poorer) fit of T$_{\rm eff} = 3450$, still much higher than the previously-adopted 2800 K temperature. Calculating M$_{\rm bol}$ for VY CMa based on these new parameters moved the star into comfortable agreement with the rightmost reaches of the 15M$_{\odot}$ evolutionary track on the H-R diagram. It is important to note that this approach was found to underestimate the luminosity of the star (Humphreys et al. 2007), This disparity could potentially be due to effects of a large-grain dust envelope producing extra "grey" extinction (Massey et al. 2008). However, a higher bolometric luminosity would still place VY CMa on the Hayashi track, in agreement with the predictions of evolutionary theory. Most recently, Choi et al. (2008) re-examined the distance to VY CMa, and report a revised value based on astrometric observations of water masers. Based on this new distance, they calculate a luminosity that is lower than previous estimates. Combined with our T$_{\rm eff}$, this brings VY CMa into very good agreement with the evolutionary tracks.

\section{Red Supergiants in the Magellanic Clouds}
We now turn our attention to the lower-metallicity environments of the Magellanic Clouds (Levesque et al.\ 2006). Massey \& Olsen (2003) show that, like the Milky Way RSGs, evolutionary tracks for the Large and Small Magellanic Clouds (LMC and SMC) do not accommodate the assumed position of RSGs on the H-R diagram. In addition, Elias et al.\ (1985) noted an interesting shift in spectral subtype when comparing RSGs in the Milky Way and Magellanic Clouds, find that the average spectral subtype shifted towards earlier types with lower metallicity. Massey \& Olsen (2003) confirmed this, measuring average RSG subtypes of M2 I, M1 I, and K5-7 I in the Milky Way, LMC, and SMC.

We obtained moderate-resolution spectrophotometry of 36 LMC RSGs and 37 SMC RSGs in November/December 2004. By fitting MARCS models of the appropriate metallicity to our spectra, as well as employing our alternate approach using the stars' ($V-K$)$_0$ colors, Levesque et al.\ (2006) determined T$_{\rm eff}$ scales for both galaxies and new bolometric luminosities for these stars. Their resulting placement on the H-R diagram brought the Magellanic Cloud RSGs into improved, though not perfect, agreement with stellar evolutionary theory. The current agreement in the LMC is good, while the agreement in the SMC is improved but still not satisfactory, with SMC RSGs showing a larger spread in T$_{\rm eff}$ across a given M$_{\rm bol}$ than their LMC counterparts. Such a spread is expected, however, due to the larger effects of rotational mixing in lower metallicity stars - for further discussion see Levesque et al.\ (2006). The agreement between the LMC and SMC RSGs and the evolutionary tracks is shown in Figure 2.

We also found two distinct explanations for the shift in spectral subtype with metallicity. Since these types are primarily based on the depths of the rich TiO bands in these spectra, which are sensitive to chemical abundances, a shift to weaker line depths and earlier spectral subtypes at lower abundances is expected, and confirmed by a comparison of our T$_{\rm eff}$ scales. Specifically, a RSG with a T$_{\rm eff}$ of 3650 K would be assigned a spectral type of M2 I in the Milky Way, M1.5 I in the LMC, and K5-M0 I in the SMC, agreeing perfectly with the shift observed by Massey \& Olsen (2003).

An additional effect is made clear when examining how the evolutionary tracks' position on the H-R diagram changes with metallicity. There is a clear shift of the Hayashi limit - the coolest tip of the tracks - to warmer temperatures at lower metallicities. For the 15-20 M$_{\odot}$ track we see a shift in the coolest T$_{\rm eff}$ for RSGs of about +100 to +150 K from the Milky Way to the LMC, and about +500 K from the Milky Way to the SMC. Following this explanation, one would expect that this shifting of the Hayashi limit should therefore impose a hard limit of how cold, and hence how late-type, RSGs can be in a particular environment. However, there are once again several unusual exceptions to this anticipated restriction that must be examined more closely.

\subsection{WOH G64}
WOH G64 is an unusual RSG in the LMC, with a number of physical properties that distinguish it from its LMC counterparts. It is surrounded by an optically thick dust torus (Elias et al.\ 1986, Roche et al.\ 1993, Ohnaka et al.\ 2008), is a known source of OH, SiO, and H$_2$O masers (Wood et al.\ 1986, van Loon et al.\ 1996, 1998, 2001, Marshall et al.\ 2004), and has nebular emission lines detected in its optical spectrum (Elias et al.\ 1986). Futhermore, observations of the optical spectra by Elias et al.\ (1986) and van Loon et al.\ (2005) show a spectrum that is dominated by very strong TiO bands, which has led to assignments of spectral types as late as M7-8, by far the latest spectral type assigned to an LMC RSG (Levesque et al.\ 2006, 2007). WOH G64 has long occupied the forbidden region to the right of the Hayashi track on the H-R diagram, in a position comparable to that of VY CMa when adopting the parameters of Le Sidaner \& Le Bertre (1996). However, recently Ohnaka et al.\ (2008) have calculated a luminosity of log($L/L_{\odot}$) = 5.45, based on modeling of the star's thick dusty torus, that is considerably lower than previous assumptions, bringing the luminosity of WOH G64 into better agreement with the rest of the LMC RSGs. Its $T_{\rm eff}$, however, remained much colder than allowed by the evolutionary tracks.

We have recently examined WOH G64 (Levesque et al.\ 2009), obtaining a spectrum and applying the MARCS stellar atmosphere models to determine $T_{\rm eff}$ for this unique RSG. Our fitting yields a $T_{\rm eff}$ of 3400 K, notably warmer than the previously-assigned 3008 K $T_{\rm eff}$ of van Loon et al.\ (2005) but still the coolest RSG T$_{\rm eff}$ in the LMC. When adopting this $T_{\rm eff}$ along with Ohnaka et al.\ (2008)'s luminosity, WOH G64 moves into much better agreement with other RSGs on the LMC H-R diagram - our currently-assumed position is shown in Figure 2. We also confirm detections of a number of emission lines in the star's optical spectrum, originally discovered by Elias et al. (1986); the lines include H$\alpha$ and H$\beta$, [O I], [N I], [S II], [N II], and [O III]. From the flux ratios of these lines, we conclude that the gas producing these lines is shock-heated and nitrogen-rich; this could potentially be similar to the nitrogen-enriched shell surrounding Sher 25 (Brandner et al.\ 1997, Smartt et al.\ 2002, Hendry et al.\ 2008).
\begin{figure}
\begin{center}
\includegraphics[scale=0.3]{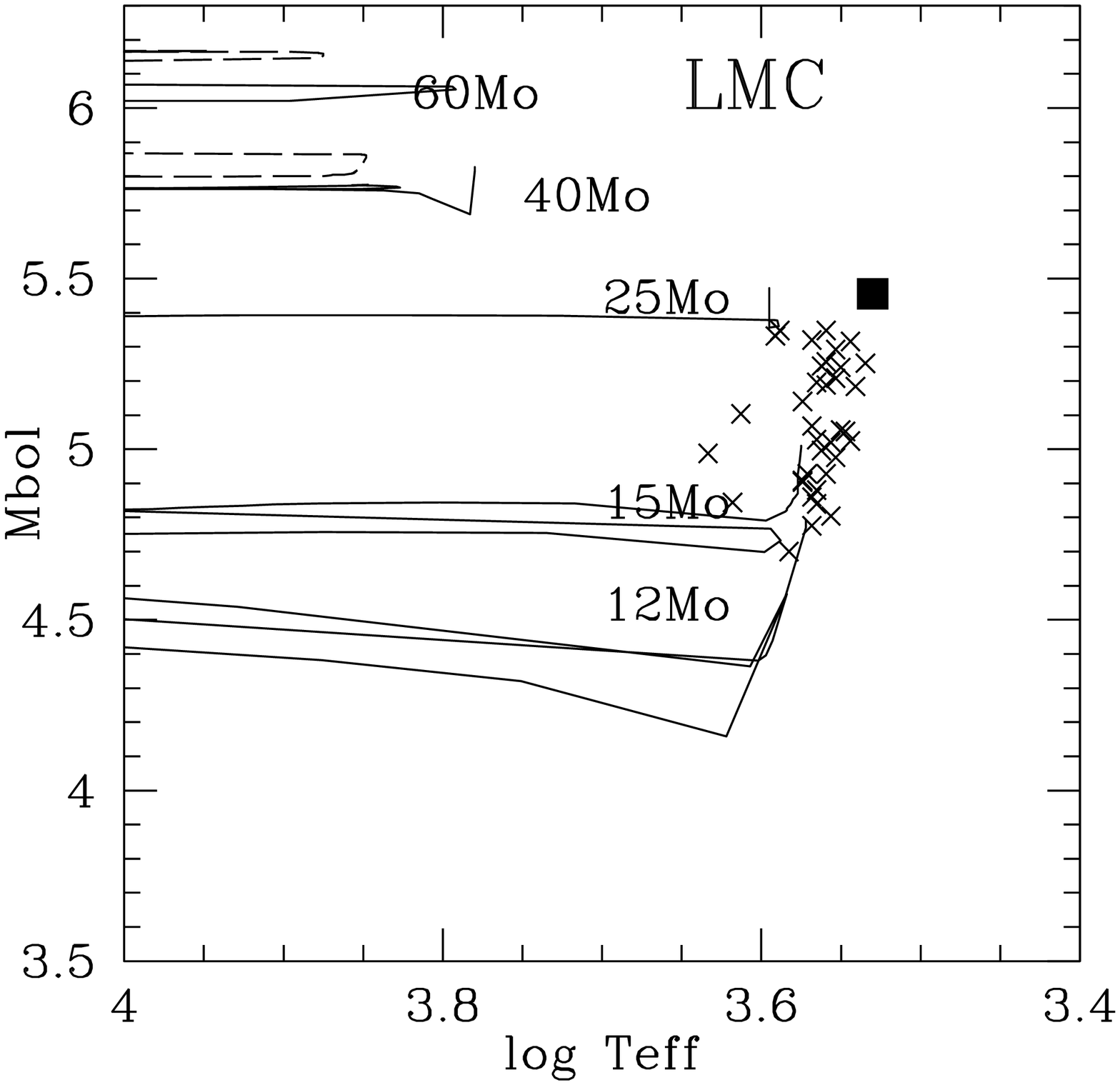}
\includegraphics[scale=0.3]{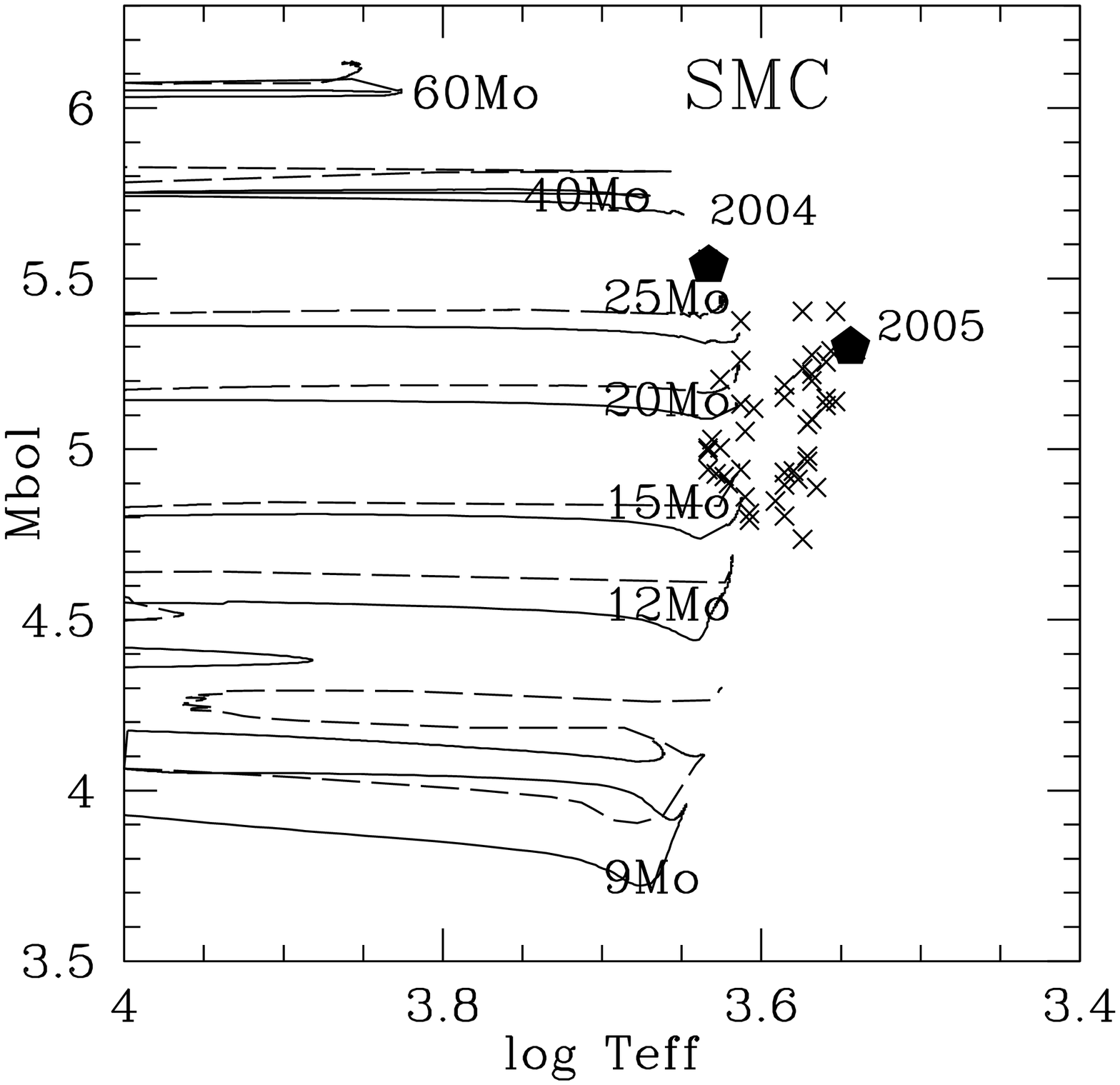}
\end{center}
\caption{Current location of the LMC (left) and SMC (right) RSGs as compared to evolutionary tracks of the appropriate metallicity, from Levesque et al.\ (2006). Non-rotating evolutionary tracks are shown as solid black lines and come from Schaerer et al. (1993b) (LMC) and Maeder \& Meynet (2001) (SMC). Tracks which adopt 300 km s$^{-1}$ initial rotation are shown as dashed lines and come from Meynet \& Maeder (2005)(LMC) and Maeder \& Meynet (2001) (SMC). WOH G64's position from Levesque et al. (2009) is shown as a filled square and compared to the LMC tracks and RSGs. HV 11423's position on the H-R diagram in both 2004 and 2005 is indicated by the filled pentagons and compared to the SMC tracks and RSGs.}
\end{figure}
\subsection{Variable Red Supergiants}
In our survey of the Magellanic Cloud RSGs, we observed several stars whose spectral subtypes, like WOH G64's, were unusually late with respect to the average type for their host galaxy. We decided to investigate these late-type RSGs further, and obtained additional spectrophotometry of seven RSGs in the LMC and five RSGs in the SMC in December 2005, a year after our original observations. When comparing these observations, we noticed large discrepancies in the spectral subtypes assigned to several of our stars in 2004 as compared to 2005. While determining spectral types involves a small degree of subjectivity, there was no comparable disagreement between the Massey \& Olsen (2003) spectral types and the ones determined for the 2004 observations in Levesque et al.\ (2006); in fact, spectral variability of a type or more is unheard of in RSGs.

The most striking example of such variation is the SMC RSG HV 11423 (Massey et al.\ 2007a). When we observed this star in 2004, it was assigned a very early spectral type of K0-1 I and a T$_{\rm eff}$ of 4300 K. When it was reobserved again in 2005, its spectrum had changed dramatically, yielding a much later spectral type of M4 I and a far cooler $T_{\rm eff}$ of 3500 K. An addition spectrum was obtained after this discrepancy was found, in September 2006; while this spectrum did not have sufficient wavelength coverage to determine a spectral type or effective temperature, it agreed excellently with the K0-1 I spectrum from 2004. The M4 I spectral type is by far the latest type we assign to an SMC supergiant, and we later uncovered an archival VLT spectrum of this star which, while not flux calibrated, clearly showed a very late spectral type of $\sim$M4.5-5. The physical parameters determined during the late-type state place HV 11423 well outside the limits of hydrostatic equilibrium - its placement in both 2004 and 2005 is included in Figure 2. In addition to strong variation in $T_{\rm eff}$, HV 11423 also displayed unusually high $V$ variability (well beyond the $\sim$1 mag variations typical of MC RSGs; Levesque et al.\ 2006), and showed variations in its $M_{\rm bol}$ and $A_V$, appearing brighter, dustier, and more luminous in its warmer state.

Most interestingly, we uncovered three more stars in our sample that exhibited the exact same behavior; two more SMC RSGs and one LMC RSG all exhibit unusual spectral and $V$ variability on the timescales of months, and vary their physical parameters in the same manner as HV 11423; when they are warmer they are brighter, dustier, and more luminous (Levesque et al.\ 2007). This variability in extinction is characteristic of the effects of circumstellar dust, and can be connected with sporadic dust production from these stars. Finally, determining $T_{\rm eff}$ for this stars with the MARCS models still leaves them at odds with stellar evolutionary theory in their cooler states, with physical parameters placing them to the right of the Hayashi track on the H-R diagram. We currently believe that these unusual properties are indicative of an unstable (and short-lived) evolutionary phase not previously associated with RSGs, and that this behavior could be due in part to the warmer limits of hydrostatic equilibrium imposed by their lower-metallicity environments.

\section{Red Supergiants in M31}
After our work in the Magellanic Clouds, we were interested in examining the properties of RSGs in M31, the most metal-rich galaxy in the Local Group (Zaritsky et al.\ 1994, Massey 2003). The Local Group Galaxies Survey (LGGS) photometry from Massey et al.\ (2006b, 2007b) allowed us to select candidates based on broad-band colors - for a complete discussion see Massey et al.\ (2009).

We obtained spectrophotometry of 16 M31 RSGs and fit these data with MARCS stellar atmosphere models computed for a 2$\times$ solar metallicity (Blair et al.\ 1982, Zaritsky et al.\ 1994). We once again derived a T$_{\rm eff}$ scale for the RSGs, although in M31 the smaller sample size precludes us from making any statistically robust conclusions about a shift in $T_{\rm eff}$ with spectral type at M31's metallicity. We do classify six M31 RSGs as M2 I, and find that the average $T_{\rm eff}$ for this subtype is 3680 K, 20 K warmer than the Milky Way M2 I subtype. This is only a marginal result, but agrees with our expectations - we find a 35 K difference between the Milky Way and LMC $T_{\rm eff}$ scales, and expect a smaller difference in $T_{\rm eff}$ between the Milky Way and M31 due to eventual saturation effects in the TiO band depths (Massey et al.\ 2009).

When placing the M31 RSGs on the H-R diagram, we originally compared their location to the 2$\times$ solar evolutionary tracks of Schaerer et al.\ (1993a) and Meynet \& Maeder (2005), and found that the higher luminosity stars were cooler than predicted; however, this turned out to be due to a simplification in the calculation of those tracks, and tracks recomputed without this approximation fit the new data very well. We note that the metallicity of M31 is not very well determined (see discussion in Crockett et al.\ 2006). Our data at present cannot address that, but observing a larger sample of RSGs to determine an average spectral subtype would also be helpful, considering this is known to shift with metallicity (Elias et al.\ 1985, Massey \& Olsen 2003, Levesque et al.\ 2006).

Finally, we used data from M31, along with the Milky Way and the MCs, to investigate Massey (1998)'s argument that RSGs are found to be more luminous at lower metallicity. We did not, in fact, confirm this trend, finding that the median luminosity of the five most luminous RSGs in each galaxy is remarkably similar, log $L/L_{\odot} \sim 5.2-5.3$ (Massey et al.\ 2009). We do not, however, consider this finding a challenge to stellar evolutionary theory, since the evolutionary tracks become nearly vertical in the RSG phase (introducing a degeneracy with luminosity and mass). Additionally, any thorough examination of RSGs and their luminosities would also require inclusion of $yellow$ supergiants, since at low metallicities the evolutionary tracks terminate at very warm temperatures (4500 K in the case of the 25M$_{\odot}$ SMC tracks; Maeder \& Meynet 2001), which correspond to G-type rather than K- or M-type supergiants (Massey et al.\ 2009).

\subsection{J004047.84+405602.6}
In addition to the 16 spectroscopicallly confirmed RSGs that we observed, we obtained a spectrum of a peculiar red star in M31, J004047.84+405602.6. This star was selected based on its extremely red color and was expected to be a M-type RSG. Instead, we were surprised to find no TiO absorption lines in the spectrum. The Ca II triplet lines are present further in the red, and allow us to confirm the membership of J004047.84+405602.6 in the M31, but despite a good signal to noise of 80 per spectral resolution element at 6000\AA, no other stellar lines are visible in the optical spectrum. With no detection of the spectral lines that would be expected for earlier type stars, including Balmer lines, this is best explained by the spectrum being heavily veiled; we suspect in this case that the absorption spectrum has been smeared out due to multiple scatterings in an expanding circumstellar dust shell (see Romanik \& Leung 1981); such scattering would affect the blue region of the spectrum more strongly than the red, perhaps explaining why the Ca II triplet is still visible in the far red (Massey et al.\ 2009).

\section{Future Work}
There is still much to be learned about RSGs, in the Milky Way as well as in other Local Group galaxies. It has been known since IRAS that RSGs have a significant dust production rate, but the driving mechanism has never been understood, as discussed by others at this meeting.  One question we plan to address observationally is whether or not the luminosity dependence of the dust production rates found by Massey et al. (2005) scale with metallicity; we can do this by analyzing Spitzer data on RSGs we've discovered in these other galaxies.  

In addition, we have turned our attention to two other Local Group galaxies with low metallicities; NGC 6822's metallicity of $Z = 0.4Z_{\odot}$ provides a middle ground between the LMC and SMC, while WLM, at $Z = 0.1Z_{\odot}$, is the lowest-metallicity Local Group galaxy currently forming stars. Using $UBVRI$ photometry from the Local Group Galaxy Survey (Massey et al.\ 2006b, 2007b) to select RSGs in both of these galaxies, we recently obtained spectra of these stars. By determining spectral types for these stars, along with physical parameters determined with the MARCS models, we will be able to construct the first detailed picture of RSGs in these galaxies and at these low metallicities.

Finally, in all of these galaxies, unusual RSGs such as VY CMa, WOH G64 require particular attention, in an attempt to understand the physical properties and evolutionary processes present in these unique and intriguing stars.

\acknowledgements
This work has been made possible through the tireless efforts of collaborators Philip Massey, Bertrand Plez, Knut Olsen, Geoff Clayton, Andre Maeder, Georges Meynet, and David Silva. We gratefully acknowledge the staff at KPNO, CTIO, the MMT, and Las Campanas for the excellent hospitality and support provided during our observations. We are grateful to John Monnier, Roberta Humphreys, George Herbig, and George Wallerstein for their correspondence regarding VY CMa; to Brian Skiff, Beverly Smith, and Roberto Mendez for helpful discussions regarding HV 11423; and Scott Kenyon for his comments on WOH G64. Philip Massey provided very helpful comments during the writing of this manuscript. Finally, many thanks to the scientific and local organizing committees of ``Hot and Cool: Bridging Gaps in Massive Star Evolution" who helped to make this meeting possible.

\end{document}